\documentclass[floatfix,aps,pra,twocolumn,longbibliography,10pt]{revtex4-1}

\usepackage{amsmath, amsthm, amssymb}

\usepackage{graphicx}

\begin{document}

\title{Femtometer Displacement Resolution with Phase-Insensitive Doppler Sensing}

\author{Bethany J. Little $^{1,3}$}
\email{bethanyjeanlittle@gmail.com}
\author{Juli\'{a}n Mart\'{i}nez-Rinc\'{o}n$^{1,3 **}$}
\thanks{The first two authors contributed equally to this work.}
\author{Umberto Bortolozzo$^{2}$}
\author{Stefania Residori$^{2}$}
\author{John C. Howell$^{1,3,4,5}$}

\affiliation{$^1$Department of Physics and Astronomy, University of Rochester, Rochester, New York 14627, USA}
\affiliation{$^2$Institut de Physique de Nice, UMR7010, Universit{\'e} de Nice-Sophia Antipolis, CNRS,1361 route des Lucioles,06560 Valbonne, France}
\affiliation{$^3$The Center for Coherence and Quantum Optics, University of Rochester, Rochester, New York 14627, USA}
\affiliation{$^4$Institute for Quantum Studies, Chapman University, 1 University Drive, Orange, California 92866, USA}
\affiliation{$^5$Racah Institute of Physics, The Hebrew University of Jerusalem, Jerusalem 91904, Israel}
\affiliation{$^{**}$Current address: Department of Physics, Stanford University, Stanford, California 94305, USA.}

 \date{February 9, 2018}
 
 \begin{abstract}
 The measurement of extremely small displacements is of utmost importance, both for fundamental studies \cite{abbott2016observation,corbitt2007all,aspelmeyer2014cavity, chan2011laser} and practical applications \cite{pierrottet2009flight,peigne2016adaptive,levi20053}.  One way to estimate a small displacement is to measure the Doppler shift generated in light reflected off an object moving with a known periodic frequency. This remote sensing technique converts a displacement measurement into a frequency measurement, and has been considerably successful \cite{kaczmarek2004laser, scalise2004self, guena2011evaluation, meier2012imaging, viza2013weak,martinez2017practical, tahmoush2015review}. The displacement sensitivity of this technique is limited by the Doppler frequency noise floor and by the velocity of the moving object.  Other primary limitations are hours of integration time \cite{viza2013weak, martinez2017practical} and optimal operation only in a narrow Doppler frequency range.  

Here we show a sensitive device capable of measuring $\mu$Hz/$\sqrt{\text{Hz}}$ Doppler frequency shifts corresponding to tens of fm displacements for a mirror oscillating at 2 Hz. While the Doppler shift measured is comparable to other techniques  \cite{viza2013weak, martinez2017practical}, the position sensitivity is orders of magnitude better, and operates over several orders of magnitude of Doppler frequency range. In addition, unlike other techniques which often rely on interferometric methods, our device is phase insensitive, making it unusually robust to noise.
 \end{abstract}

\maketitle

Remote sensing, which may be defined as the acquisition of information about an object without physical contact \cite{elachi2006introduction}, has existed since humans first looked at objects. However the term, which first gained popularity in the 1960's \cite{campbell2011introduction}, has been primarily used in the fields of geology and astronomy to measure relatively large scale information; Doppler measures of moving humans or birds are considered small on these scales \cite{tahmoush2015review, chen2003micro}. 

The small-displacement limits of sensing have been pushed with various reported methods.  By measuring the beat frequency of a laser beam reflected of a mirror oscillating at 10 kHz, a noise equivalent displacement of 10 fm/$\sqrt{\text{Hz}}$ after 10 s of integration time was reported \cite{weksler1980measurement}, and similar results have been achieved with optomechanical approaches \cite{aspelmeyer2014cavity}.   Weak-values techniques have been successful for velocity measurements in the low frequency regime of a Michelson interferometer \cite{viza2013weak, martinez2017practical}, reporting 60 fm/s for an oscillating mirror frequency of 2.5 mHz and an integration time of about 40 hours \cite{martinez2017practical}.  Non-classical light has been used to reach the ~fm/$\sqrt{\text{Hz}}$ regime in MEMS cantilevers \cite{pooser2015ultrasensitive}.  On a much larger oscillator mass scale, the LIGO collaboration reached a remarkable noise floor of about $10^{-20}$ m/$\sqrt{\text{Hz}}$ for mirror oscillations of 100 Hz \cite{abbott2016observation}.  With a position sensitivity spanning the range from nanometers down to tens of femtometers, and operating in an oscillator frequency regime of single Hz, this paper demonstrates a device that is both sensitive and versatile enough to place remote sensing in a new realm of ultra-small displacement measurements.

The key to the success of the experiment described here is the remarkable combination of frequency sensitivity and the phase insensitivity arising from two-wave mixing inside a highly dispersive liquid crystal light valve (LCLV) \cite{residori2008slow, bortolozzo2010slow}. 
In a previous work, Bortolozzo et al. \cite{Bortolozzo:13} demonstrated that the linear dispersive regime of the LCLV can be used to measure doppler shifts down to a microhertz; here we extend this work.  
  By using a periodic oscillating mirror movement, we are able to determine the minimum position sensitivity of the device.  Using measurements of up to ten minutes, we also measure the noise spectrum of the sensor in the low frequency regime (sub-Hertz), demonstrating remarkable stability and sensitivity in a notoriously noisy frequency regime.

	\begin{figure}
	\includegraphics[width=.8\columnwidth]{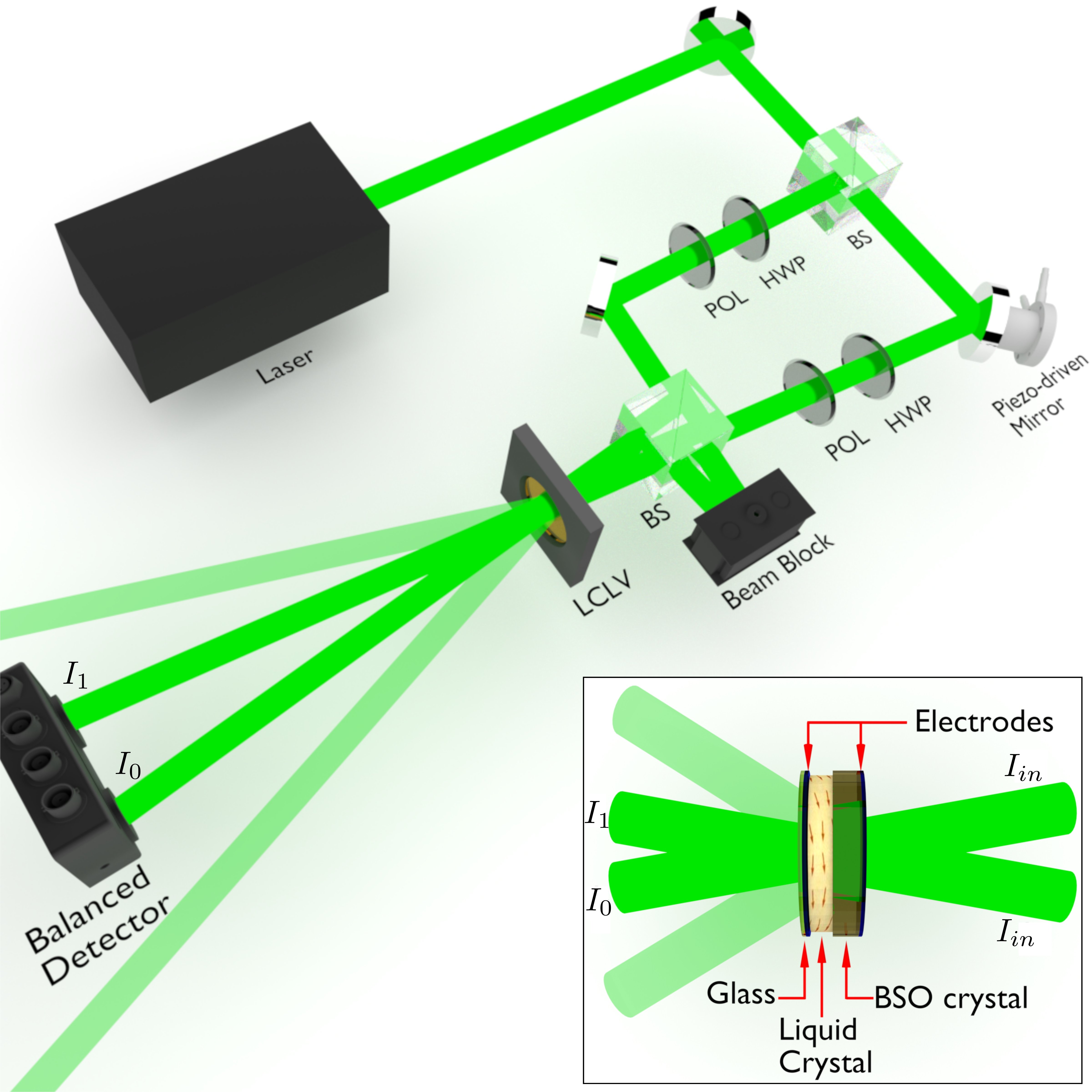}
	\caption{Diagram (not to scale) of the experimental setup. A laser of wavelength 532 nm is divided by a beamsplitter (BS). One beam is Doppler shifted by reflecting off a moving piezo-driven mirror, and is incident with the unshifted beam on the LCLV. The first-order output beams with intensities $I_0$ and $I_{1}$ are incident on a balanced detector, and the sum and difference are measured. Polarizer (POL) and half-wave plate (HWP) combinations control the power and polarization of each beam.  The inset shows the components of the LCLV. A photorefractive layer (BSO) on the entrance face and a glass plate on exit face hold in place the liquid crystal layer. Transparent electrodes are used to apply an external bias voltage which enables tuning of the response of the device. The angle of 0.3 degrees between the two beams incident on the LCLV is exaggerated. A combination of lenses and irises (not shown) are used to focus the first orders onto the detector and filter out the higher orders produced by the wave mixing int the LCLV.}\label{fig:exp}
	\end{figure}

In our setup, shown in Figure \ref{fig:exp}, a moving piezo-driven mirror generates a Doppler shift in one beam relative to a reference beam, after which both beams interact in the LCLV.  Since the LCLV is highly sensitive to small frequency shifts, we can use it to measure Doppler shifts caused by extremely small displacements of the mirror \cite{bortolozzo2013transmissive}.

The Doppler shift of light incident at 45$^\circ$ on a mirror moving with a velocity $v_m$ is given by 
 $f_d = \sqrt 2 v_m/\lambda$, where $\lambda$ is the wavelength of the light. If the mirror follows a harmonic oscillation with amplitude A and frequency $f_m$, then the maximum Doppler shift induced in the reflected light is
	\begin{equation}\label{eq:Doppler}
	 f_d = \sqrt 2 k A f_m,
	\end{equation}
where $k=2\pi/\lambda$ is the wavenumber. 

We now give a brief description of the LCLV and its expected frequency response. The LCLV derives its unique properties from the combination of a liquid crystal layer and a photorefractive layer. The photorefractive layer is an inorganic bismuth silicate crystal (Bi$_{12}$SiO$_{20}$), transparent in the visible range. This acts as one side of a cell wall for the liquid crystal, which is held in place on the other side by a glass (BK7) window, as shown in the inset of Figure \ref{fig:exp}. Spacers a few microns long define the thickness of the liquid crystal. Transparent electrodes placed within the crystal sandwich allow tuning of the LCLV's properties by means of an applied AC voltage. 

The index of refraction of the liquid crystal varies dramatically depending on the orientation of its long molecules, which rotate according to the potential difference between the plates. Light incident on the bismuth silicate crystal creates an intensity dependent spatial modification of the effective voltage across the liquid crystal layer. This causes the molecules of the liquid crystal to rotate locally, creating a spatially dependent index \cite{Bortolozzo2008}.

When two beams of equal intensity $I_{in}$ and wave vectors $\vec{k_1}$ and $\vec{k_2}$ are incident on the device, their combined field creates a grating within the LCLV, with grating wave vector $\vec{K_g} = \vec{k_1}-\vec{k_2}$. The grating allows energy exchange between the beams; the interaction may be viewed as a wave mixing process. If the periodicity of their interference, $\Lambda \equiv 2 \pi/K_g$, is much greater than the thickness $d$ of the liquid crystal, the diffracted orders of the output can be characterized according to nonlinear wave mixing in the Raman-Nath regime \cite{Bortolozzo2008}.  This two-beam process has been shown to have remarkably slow group velocities \cite{residori2008slow}, and corresponding group delays $\tau$ on the order of hundreds of milliseconds over only a few microns of material.
 
The $m^{th}$ order output intensity of the wave mixing is given by \cite{Bortolozzo:2009dr}
\begin{equation}
\frac{I_m}{ I_{in}} = | J_{m-1}(\rho )+ i J_{m}(\rho) e^{-i\Psi}|^2 
\end{equation}

where $\tan \Psi = f_d \tau/(1+l_d^2K_g^2)^2$, and $J_m$ denote the Bessel functions of the first kind, with argument
\begin{equation}
\rho = \frac{2 k n_2 I_{in} d}{\sqrt{(1 + l_d^2 K_g^2)^2 + (f_d \tau )^2}}.
\end{equation}
 Here $n_2$ is the equivalent Kerr-type nonlinear index, and $l_d$ is the transverse diffusion length, determined by the material constants and the applied AC voltage \cite{prost1995physics} (See Methods).  The normalized intensity difference between the zeroth and first-order beams is a function of the frequency difference between them and is given by
\begin{equation}\label{eq:DN}
\frac{\Delta I}{I_{in}}=\frac{ I_1(f_d)- I_{0}(f_d)}{I_{in}}
\end{equation}

For small Doppler frequency shifts, there is a linear response,
\begin{equation}\label{eq:linres}
 \Delta I \approx f_d \chi I_{in},
\end{equation}
the slope of which can be calculated as \cite{Bortolozzo:13} 
\begin{equation}\label{eq:chi}
\chi = 8 \pi \tau J_0(2 k d n_2 I_{in})J_1(2 k d n_2 I_{in}).
\end{equation}

The value of $\chi$ is dependent on a variety of experimental parameters, including the group index, which is related to $\tau$, the nonlinear Kerr coefficient $n_2$, and the intensities of the beams incident on the LCLV.  For our experiment, $\chi$ is estimated to be 0.5 s (see Methods).

We note that, unlike common interferometric methods for measuring small frequency shifts and displacements, the amplitude of our signal does not depend on the relative phase between the two beams. This is similar to other wave mixing processes, such as inside acousto-optic modulators. The large linear response $\chi$ and this phase insensitivity are the major advantages of using a LCLV.

The experimental setup is shown in Figure \ref{fig:exp}.  Collimated light from a laser at 532 nm is split on a 50/50 beamsplitter, and one beam reflects off a piezo-driven (PI S310) mirror and experiences a Doppler shift relative to the other beam. A Mach-Zehnder type setup is used to create a slight angle between the two beams on the order of 0.3 degrees, and both beams are then incident on the LCLV.  Note that the two beams do not recombine on the beamsplitter before the LCLV.  The small angle between the beams ensures that the interaction occurs in the Raman-Nath regime \cite{nath1938diffraction, moharam1978criterion,Bortolozzo2008}. The two primary diffracted output orders of the LCLV are focused onto a balance detector (Thorlabs PDB210A2/M). Irises are used to block light from the higher diffracted orders. 

\begin{figure}
	\includegraphics[width=\columnwidth]{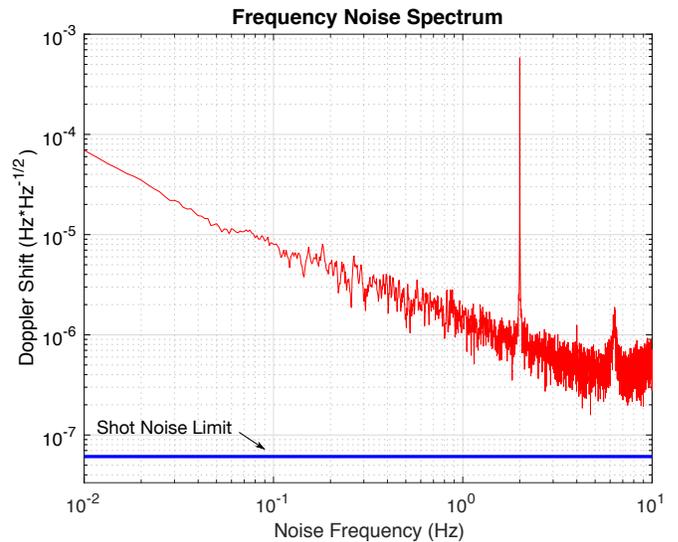}
	\caption{Noise spectrum for the device, with a 100 mV reference signal at 2 Hz applied to the piezo stack. We obtained the spectrum by taking a FFT of a 10 minute scan. The reference signal, which corresponds to a Doppler shift of 200 mHz, is used to normalize the vertical axis.  The spectrum is shown up to the bandwidth of the linear region of the LCLV, around 10 Hz.  The blue line is the SNL based on the estimated value of $\chi$.  At 2 Hz, the noise floor is around 1 $\mu\text{Hz}/\sqrt{\text{Hz}}$; for a mirror oscillating at 2 Hz, this corresponds to a displacement sensitivity of around 20 fm/$\sqrt{\text{Hz}}$.}\label{fig:noisespec}
	\end{figure}

In order to minimize noise due to turbulence, temperature drifts, and room light, the apparatus is enclosed inside an insulating foam box. The optical table is not floated and no further isolation measures are used, an indication of the remarkable stability of this technique. For small Doppler shifts, a lock-in amplifier is used to measure the signal from the balance detector, which is fed through a high pass filter to avoid overload of the lock-in due to DC drifts.  For measurement of the noise spectrum (Figure \ref{fig:noisespec}) no filtering is used.

The LCLV used has a thickness $d=9$ $\mu$m, a group delay $\tau = 72$ ms, and is optimized by a 2.6 V AC driving voltage at 1kHz, giving an estimated effective nonlinear index $n_2 =- 1.8 \times 10^{-4}$ $W/m^2$, $l_d \approx$ 12 $\mu$m, and linear response $\chi = 0.5$ s.  The beams at the LCLV have a 1/$e^2$ diameter of 4 mm, and the interference spacing is $\Lambda \approx 110$ $\mu$m.  The mirror is modulated at a fixed frequency of 2 Hz, with an amplitude determined by the applied voltage and the manual-specified piezo response of 60 nm/V.

	
	\begin{figure}
	\includegraphics[width=\columnwidth]{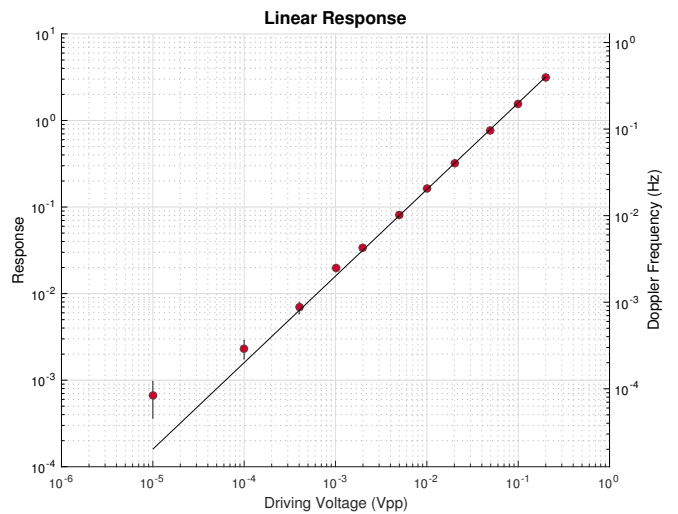}
	\caption{The difference signal at the detector is shown as a function of the peak to peak amplitude of the the voltage applied to the piezo.  The response is linear above $10^{-4}$ V. The black line is the theoretical linear response.  At low voltages, the electronics of the piezo driver become nonlinear and the applied voltage from the driver is no longer equal to the voltage on the piezo.}\label{fig:results}
	\end{figure}

The results of the device depend on both the noise floor and the linearity of the response. We analyze both of these criteria, and follow with  a discussion of the range and practicality of the device. 

The fundamental noise limit for laser power measurements using a coherent source is given by the shot noise $\sqrt N$.  Using equation \eqref{eq:linres}, and writing the power $P$ in terms of the photon number $N$ by using the energy per photon, $hc/\lambda$, we find the limit for measuring a frequency shift with this device,
	\begin{equation}\label{eq:fSN}
	f_{SN}=\frac{1}{\chi}\sqrt{\frac{1}{N}}=\frac{1}{\chi}\sqrt{\frac{hc}{\eta \lambda P T}},
	\end{equation}
where $\eta$ is the detector efficiency and $T$ is the acquisition time per measurement.

The limit of Equation \eqref{eq:fSN} is plotted in Figure \ref{fig:noisespec} along with the measured noise spectrum for the device when the mirror is driven at 2 Hz. We normalize the spectrum using a reference signal of 100mV on the piezo, which gives a Doppler shift of 200 mHz.  
The spectrum was obtained by taking the finite Fourier transform of a 10 minute scan, with no electrical or frequency filtering. The power incident on the detector is 1.15 mW, and the detector efficiency is $\eta = 0.35$ giving, for our experimental parameters, a shot noise limit of $f_{SN} $= 60 nHz/$\sqrt{\text{Hz}}$.

In order to obtain the normalization of Figure \ref{fig:noisespec}, the reference signal must be well within the linear response region of the device.  Figure \ref{fig:results} shows the measured response of the LCLV for different applied Doppler shifts, generated by a signal at 2 Hz and with varying peak-to-peak voltages. The linear response is evident for voltages above $10^{-4}$ $V_{pp}$.  The nonlinear response of the electronics driving the piezo limit the application of smaller voltages, and result in divergence from the expected linear behavior.  The reference signal at 200 mHz, however, is well within the linear region.  The data in Figure \ref{fig:chi} in the Methods Section also confirms that the reference signal is within the linear response region. 

We find the shot noise limited displacement using Equations \eqref{eq:Doppler} and \eqref{eq:fSN}:
	\begin{equation}\label{eq:ASN}
	A_{SN}=\frac{1}{ \sqrt{2} k f_m }f_{SN}=\frac{1}{ k f_m \chi} \sqrt{\frac{hc}{2\eta \lambda P T}}.
	\end{equation}

\begin{figure}
	\includegraphics[width=\columnwidth]{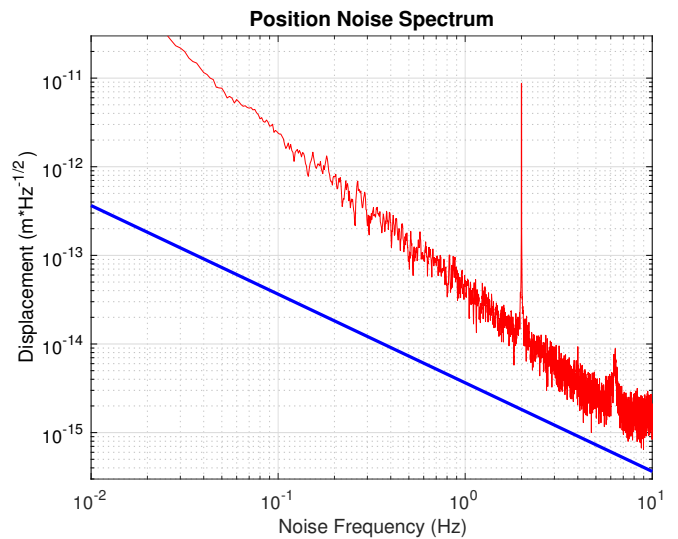}
	\caption{The noise spectrum for position measurement, calculated from the spectra of Figure \ref{fig:noisespec}, is plotted.  The blue line is the shot noise-limited sensitivity for position measurement from Equation \eqref{eq:sens}.  For a mirror frequency of 2 Hz, the noise is around 20 fm/$\sqrt{\text{Hz}}$, which is only about an order of magnitude above the calculated shot noise limit of 1.8 fm/$\sqrt{\text{Hz}}$ at that frequency.
 }\label{fig:noisepos}
	\end{figure}

 Since the displacement of the mirror can be calculated as a function of the measured Doppler shift according to Eq. \eqref{eq:Doppler}, the sensitivity of the device is correspondingly proportional:
	\begin{equation}\label{eq:sens}
		\delta A =\frac{1}{ \sqrt{2} k } \frac{ \delta f_d(f_m)}{ f_m}.
	\end{equation}
Here $\delta A $ is the position sensitivity, and $\delta f_d$ the Doppler sensitivity.  This relationship is used to plot the noise spectrum for position in Figure \ref{fig:noisepos}. For our experimental parameters, $A_{SN}$= 1.8 fm/$\sqrt{\text{Hz}}$. Figure \ref{fig:noisespec} shows a Doppler sensitivity at our reference mirror frequency of 2 Hz to be around 1 $\mu\text{Hz}/\sqrt{\text{Hz}}$.  The corresponding displacement noise floor, shown in Figure \ref{fig:noisepos} is calculated to be around 20 fm/$\sqrt{\text{Hz}}$, which is only about an order of magnitude above the shot noise limit.


To better compare our results to other Doppler measures, we may also consider the velocity of the moving mirror, given by $v_m = f_d \lambda/\sqrt 2$.  Considering the measured noise floor in Figure \ref{fig:noisespec} to be $\mu$Hz$/\sqrt{\text{Hz}}$, the corresponding measurable velocity is 400 fm/s/$\sqrt{\text{Hz}}$. This is similar to the result found in Ref \cite{viza2013weak}; however in their case the measurement was achieved by averaging about two hours worth of data; in ours this is the sensitivity per $\sqrt{\text{Hz}}$.  Even one hour of integration time with our device gives, according to equation \eqref{eq:fSN}, a shot-noise limited velocity measurement of 6 fm/s.  Using a combination of balance detection and weak values, sensitivities as small as 60 fm/s have recently been reported \cite{martinez2017practical}; however this was only after 40 hours of integration time. Thus we see that not only is our technique comparable to other precision velocity measurements, but it is also vastly more robust.

We have proposed and demonstrated a remote sensing technique which makes use of a LCLV to measure Doppler shifts and corresponding displacements with a noise floor on the order of 20 fm/$\sqrt{\text{Hz}}$ in the Hz regime.  Direct measurement of displacements this small were limited not by the device, but by the nonlinearity of the piezo and driver for low voltages.  In principal, without this added noise in the moving object, the LCLV measuring device is linear down to arbitrarily small measurements, and is limited only by the noise floor of the measurement. The Doppler frequency noise floor of the technique, measured to be around $\mu$Hz/$\sqrt{\text{Hz}}$, is only about an order of magnitude above the shot noise limit, remarkable for a tabletop experiment with no isolation, damping, or filtering. 

Many other measurement techniques operate in a higher frequency regime, where it is not only easier to obtain higher precision by averaging, but also typically easier to approach the shot noise limit. In contrast, our device is optimized to work at 2 Hz, a frequency regime of great interest, for instance to the gravitational-wave community \cite{amaro2012low}. This optimal frequency range of the device is set primarily by the thickness of LCLV.  We plan to investigate working at even lower frequencies by tuning the thickness and driving voltages of LCLVS to optimize their behavior at low frequency. 

 We believe this new regime of remote sensing will have many novel applications.  For example, we anticipate the LCLV could be used for micro-mechanical cooling applications where it has proved difficult to measure the ground state.  The range could be useful in biological applications in bridging the gap toward increasingly smaller scales.  Finally, in a return to the original motivation for the development of liquid crystals, and in line with the history of remote sensing, we plan to explore the femtometer scale imaging capabilities of the LCLV.

\section{Acknowledgements and Funding}

This work is supported by the Direction G{\'e}n{\'e}rale de l'Armement (DGA), under the contract ANR-14-ASMA-0004-01, Astrid Maturation HYDRE, the Army Research Office (ARO) (W911NF-12-1-0263), and Northrop Grumman Corporation.

\section{Methods}
\subsection{Determination of $\chi$}
	\begin{figure}
	\includegraphics[width=\columnwidth]{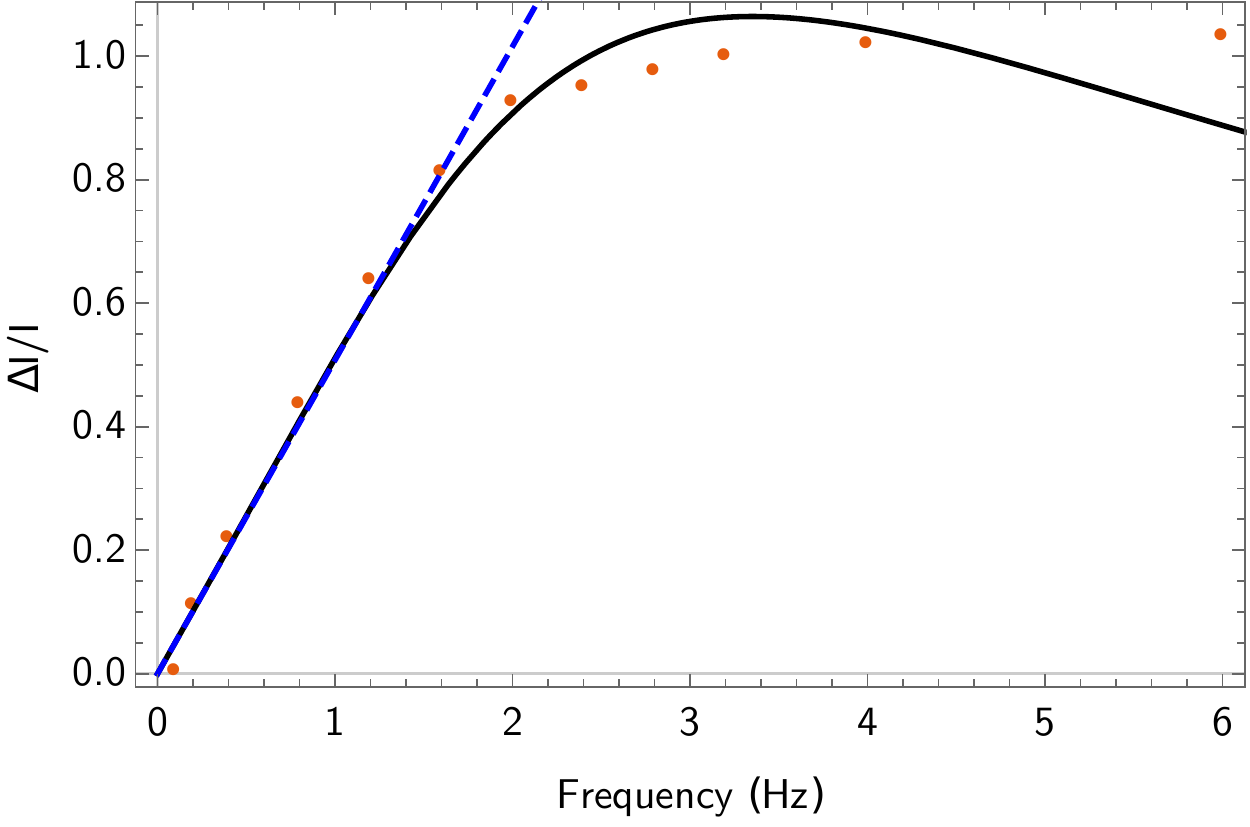}
	\caption{Data (red dots) and fit (black) used to estimate the value of the slope $\chi$ of the linear region (see dashed line). The fit yields $\chi = 0.5$ s.}\label{fig:chi}
	\end{figure}
An estimate of $\chi$ may be obtained by generating increasingly large frequency shifts between the two beams to determine the end of the linear region of the response curve.  Fitting this response to the full expression for the expected response allows determination of $n_2$ and calculation of $\chi$ using Equation \eqref{eq:chi}.  Figure \ref{fig:chi} shows the data (red circles) and theory fit for this calculation, as well as the slope $\chi=0.5$ s of the linear region (blue, dashed line).

 \subsection{Diffusion Length}
The transverse diffusion length is
\begin{equation}
 l_d=\frac{d\sqrt{\Delta_\epsilon / K}}{V},
\end{equation}
where $\Delta_\epsilon$ is the dielectric anisotropy, $K$ is the elastic
constant of the liquid crystal, V is the AC voltage applied to the liquid crystal, and $d$ is the thickness of the cell.

The diffusion length $l_d$ is determined by observing the point spread function. Light is tightly focused on the BSO crystal, which induces a refractive index variation $\delta n$ of the same shape as the light spot, but of different transverse size; this transverse index variation gives $l_d$. For our cell, $l_d\approx$ 12 microns.

\bibliography{LittleArxiv}

\end{document}